\RequirePackage[british]{babel}
\documentclass[reqno,a4paper,12pt,final]{amsart}
\usepackage[utf8x]{inputenc}
\usepackage[a4paper,hmargin=2cm,tmargin=3cm,bmargin=3cm]{geometry}
\usepackage{amsmath,amssymb,amstext,amsthm,amscd,mathrsfs,eucal}
\usepackage{graphicx,color}
\usepackage{array}
\usepackage[notref,notcite]{showkeys}
\usepackage{hyperref}
\hypersetup{%
  pdftitle   = {Simplicity in the Faulkner construction},
  pdfkeywords = {M2, M Theory, Bagger-Lambert, Lie, Faulkner, Lie triple system},
  pdfauthor  = {José Figueroa-O'Farrill},
  pdfcreator = {\LaTeX\ with package \flqq hyperref\frqq}
}
\PrerenderUnicode{éÉ}
\let\into\hookrightarrow

\newcommand{\fg}{\mathfrak{g}}

\newcommand{\fk}{\mathfrak{k}}

\newcommand{\CC}{\mathbb{C}}
\newcommand{\ZZ}{\mathbb{Z}}
\newcommand{\XX}{\mathbb{X}}
\newcommand{\eD}{\mathscr{D}}

\DeclareMathOperator{\im}{im}

\newcommand{\hfl}[1]{\overline{#1}^\flat}

\theoremstyle{plain}
\newtheorem{lemma}{Lemma}

\newtheorem{theorem}[lemma]{Theorem}

\theoremstyle{definition}

\newcommand{\MUNCH}[1]{\relax}

\allowdisplaybreaks
\begin{document}
\title{Simplicity in the Faulkner construction}
\author{José Figueroa-O'Farrill}
\address{Institute for the Physics and Mathematics of the Universe, University of Tokyo, Kashiwa, Chiba 277-8586, Japan}
\address{School of Mathematics and Maxwell Institute for Mathematical Sciences, University of Edinburgh, Edinburgh EH9 3JZ, UK}
\email{Jose.Figueroa@ipmu.jp}
\date{\today}
\begin{abstract}
  We revisit the Faulkner construction of metric 3-Leibniz algebras admitting an embedding Lie (super)algebra.  In the case of positive-definite signature, we relate the various notions of simplicity: of the 3-algebra, of the representation and of the embedding Lie (super)algebra.  This reduces their classification to the extant classifications of simple Lie (super)algebras.
\end{abstract}
\maketitle
\tableofcontents

\section{Introduction}
\label{sec:introduction}

Starting with the pioneering proposal of Bagger and Lambert \cite{BL1,BL2} and Gustavsson \cite{GustavssonAlgM2} for a nonabelian superconformal field theory of M2-branes, certain kinds of ternary algebras (contained in the class of \emph{metric 3-Leibniz algebras}, defined for example in \cite{JMF3lectures}) are known to be associated to three-dimensional superconformal Chern--Simons + matter theories with various amounts of supersymmetry.  The case of maximal ($N{=}8$) supersymmetry corresponds to the metric 3-Lie algebras of Filippov \cite{Filippov}, whereas for less-than-maximal supersymmetry a number of algebras have been shown to play a rôle, starting with the work of Bagger and Lambert \cite{BL4} for the $N{=}6$ theories of Aharony, Bergman, Jafferis and Maldacena \cite{MaldacenaBL} and Cherkis and Sämann \cite{CherSaem} for $N{=}2$ theories.  In \cite{Lie3Algs} a construction was given of all the 3-algebras underlying superconformal Chern--Simons + matter theories in three dimensions, purely in terms of Lie algebras and representation theory, by specialising a construction due to Faulkner \cite{FaulknerIdeals}.  The mathematical literature is replete with a bewildering array of triple systems, to which some authors have linked the 3-algebras of interest \cite{NilssonPalmkvist,ChowJTS,Palmkvist}.

In particular, in a recent paper \cite{Palmkvist} Palmkvist revisits the relation between 3-algebras of the $N{=}6$ theories and Lie superalgebras established in \cite{Lie3Algs}, refining the result by showing that, in the case of positive-definite signature, several notions of simplicity (or irreducibility) correspond.  This is phrased---in the opinion of this author, unnecessarily---in the language of generalised Jordan triple systems and the purpose of this note is to show that this refinement can be recovered easily from the construction in \cite{Lie3Algs}.  In fact, as we will see, the proof that the simplicity of the triple system corresponds to that of the embedding Lie (super)algebra is most easily obtained by first showing that both are separately equivalent to the irreducibility of the representation in the Faulkner construction.  In other words, it is at the end of the day a result in representation theory, which is precisely the language in which the Faulkner construction allows us to phrase properties of the 3-algebras.  For completeness we also treat the cases of Lie and anti-Lie triple systems, as a special case of the real and quaternionic Faulkner constructions, respectively.  The latter case corresponds to the triple systems underlying (at least some of) the $N{=}4,5$ Chern--Simons + matter theories.

This note is organised as follows.  In §~\ref{sec:faulkner} we briefly recall the main definitions of the Faulkner construction of metric 3-Leibniz algebras: paying particular attention to the real orthogonal, complex unitary and quaternionic unitary representations.  In §~\ref{sec:simplicity} we consider the special cases of the above constructions when the triple system admits an embedding into a Lie (super)algebra.  In those cases there are three notions of simplicity (or irreducibility): simplicity of the triple system, irreducibility of the representation and simplicity of the embedding Lie (super)algebra and we show how they are related.  We show that in the case of positive-signature, the three notions are essentially the same.  The precise statements are contained in Theorem \ref{thm:LTS-simplicity} for the Lie triple systems, Theorem \ref{thm:BL4-simplicity} for the $N{=}6$ triple systems and Theorem \ref{thm:aLTS-simplicity} for the anti-Lie triple systems arising from a quaternionic unitary representation.

\section{The Faulkner construction}
\label{sec:faulkner}

In this section we review the Faulkner construction of 3-algebras in \cite{Lie3Algs}.

Let $\fg$ be a real finite-dimensional Lie algebra with an ad-invariant symmetric bilinear form $\left(-,-\right)$ and let $V$ be a finite-dimensional faithful representation of $\fg$ with dual representation $V^*$.  We will let $\left<-,-\right>$ denote the dual pairing between $V$ and $V^*$.  Transposing the $\fg$-action defines for all $v\in V$ and $\alpha \in V^*$ an element $\eD(v, \alpha)\in \fg$ by
\begin{equation}
  \label{eq:D-map}
  \left(X,\eD(v, \alpha)\right) = \left<X \cdot v, \alpha\right> \quad\text{for all $X\in\fg$,}
\end{equation}
where the $\cdot$ indicates the $\fg$-action on $V$.  Extending $\eD$ bilinearly, defines a $\fg$-equivariant map $\eD : V \otimes V^* \to \fg$, which as shown in \cite{Lie3Algs} is surjective because $V$ is faithful.  The $\fg$-equivariance of $\eD$ is equivalent to the \emph{fundamental identity}
\begin{equation}
  \label{eq:FI-Faulkner}
  [\eD(v,\alpha),\eD(w,\beta)] =  \eD(\eD(v,\alpha)\cdot w ,\beta) + \eD(w, \eD(v,\alpha)\cdot \beta)~,
\end{equation}
for all $v,w \in V$ and $\alpha ,\beta \in V^*$, where the dual action $\eD(v,\alpha)\cdot \beta$ is defined by
\begin{equation}
  \label{eq:dualmodule}
  \left<w, \eD(v,\alpha)\cdot \beta\right> = -\left<\eD(v,\alpha)\cdot w, \beta\right>~.
\end{equation}
The map $\eD$ defines in turn a trilinear product
\begin{equation}
  \label{eq:3-product}
  \begin{aligned}[m]
    V \times V^* \times V &\to V\\
    (v,\alpha,w) &\mapsto \eD(v,\alpha) \cdot w~.
  \end{aligned}
\end{equation}

A special case of the Faulkner construction is where $V$ is a faithful unitary representation of $\fg$.  This means that $V$ is a real, complex or quaternionic representation of $\fg$ possessing a $\fg$-invariant real symmetric, complex hermitian or quaternionic hermitian inner product, respectively.  This gives rise, respectively, to a real orthogonal, complex unitary or quaternionic unitary representation of $\fg$.  As shown in \cite{Lie3Algs}, the real case corresponds precisely to the metric 3-Leibniz algebras constructed by Cherkis and Sämann in \cite{CherSaem}, whereas the complex case contains as a special class the $N{=}6$ 3-algebras of \cite{BL4}.

\subsection{The real Faulkner construction}
\label{sec:real}

Let us first consider the case of $V,\left<-,-\right>$ a real inner product space admitting a faithful orthogonal action of a real metric Lie algebra $\fg, \left(-,-\right)$.  The inner product on $V$ sets up an isomorphism $\flat: V \to V^*$ of $\fg$-modules, defined by $v^\flat = \left<v,-\right>$, with inverse $\sharp: V^* \to V$.  The map $\eD: V \otimes V^* \to \fg$ defined by equation \eqref{eq:D-map} induces a map $D: V \otimes V \to \fg$, by $D(v, w) = \eD(v, w^\flat)$.  In other words, for all $v,w\in V$ and $X \in \fg$, we have
\begin{equation}
  \label{eq:D-map-R}
  \left(D(v, w),X\right) = \left<X \cdot v, w\right>~.
\end{equation}
It follows from the $\fg$-invariance of the inner product that
\begin{equation}
  \left(D(v,w),X\right)  = \left<X \cdot v, w\right> = - \left<X \cdot w, v\right>  = - \left(D(w,v),X\right)~,
\end{equation}
whence
\begin{equation}
  \label{eq:D-skew}
  D(v, w) = - D(w, v)~.
\end{equation}

Using $D$ we can define a 3-bracket on $V$ by
\begin{equation}
  \label{eq:3-bracket-R}
  [u,v,w] := D(u, v) \cdot w~,
\end{equation}
for all $u,v,w \in V$.  The resulting 3-Leibniz algebra, which appeared originally in \cite{FaulknerIdeals} but more recently in
\cite{CherSaem} in the context of superconformal Chern--Simons + matter theories, satisfies the following axioms for all $x,y,z,v,w\in V$:
\begin{enumerate}
\renewcommand{\labelenumi}{(\alph{enumi})}
\item the \emph{orthogonality} condition
  \begin{equation}
    \label{eq:unitarity-R}
    \left<[x,y,z], w\right> =  - \left<z, [x,y,w]\right>~;
  \end{equation}
\item the \emph{symmetry} condition
  \begin{equation}
    \label{eq:symmetry-R}
    \left<[x,y,z], w\right> = \left<[z,w,x], y\right>~;
  \end{equation}
\item and the \emph{fundamental identity}
  \begin{equation}
    \label{eq:FI-R}
    [x,y,[v,w,z]] - [v,w,[x,y,z]] = [[x,y,v],w,z] + [v,[x,y,w],z]~.
  \end{equation}
\end{enumerate}
It follows from the orthogonality and symmetry conditions that $[x,y,z] = - [y,x,z]$ for all $x,y,z \in W$, which is nothing but equation \eqref{eq:D-skew}.  A special case are the algebras appearing in the maximally supersymmetric $N=8$ theory of Bagger--Lambert \cite{BL1,BL2} and Gustavsson \cite{GustavssonAlgM2}, for which the 3-bracket is totally skewsymmetric.  Another special case of these 3-Leibniz algebras corresponds to metric Lie triple systems, for which the 3-bracket obeys $[x,y,z] + [y,z,x] + [z,x,y] =0$.  Metric Lie triple systems are characterised by the fact that they embed into $\fg \oplus V$ as a real metric $\ZZ_2$-graded Lie algebra and are in one-to-one correspondence with pseudoriemannian symmetric spaces.

\subsection{The complex Faulkner construction}
\label{sec:complex}

Let $V,h$ be a complex unitary representation of a real metric Lie algebra $\fg,\left(-,-\right)$.  In our somewhat odd conventions, the hermitian inner product $h$ is complex linear in the left entry and complex antilinear in the right one.  The hermitian inner product $h$ sets up a complex antilinear isomorphism $V \to V^*$.  Indeed, we may parametrise $V^*$ by elements of $V$ via $v \mapsto \hfl{v} := h(-,v)$.  The Faulkner map \eqref{eq:D-map} now defines a sesquilinear map $D: V \times V \to \fg_\CC$ to the complexification of $\fg$, defined by 
\begin{equation}
  \label{eq:D-map-C}
  \left(D(v, w),\XX\right) = h(\XX \cdot v, w)~,
\end{equation}
for all $v,w\in V$ and $\XX \in \fg_\CC$.  We have extended the action of $\fg$ on $V$ to $\fg_\CC$ in the naive way: $(X + i Y) \cdot v = X\cdot v + i Y \cdot v$, and we have extended the inner product on $\fg$ to $\fg_\CC$ complex \emph{bilinearly}, turning $\fg_\CC$ into a complex metric Lie algebra acting faithfully on $V$.  As shown in \cite{Lie3Algs}, this means that $D$ is surjective.  Of course, being complex, $\fg_\CC$ cannot preserve $h$ --- instead it obeys $h(\XX \cdot v, w) = - h(v, \overline\XX \cdot w)$.  Complex conjugating \eqref{eq:D-map-C}, we see that this means that $\overline{D(u,v)} = - D(v,u)$.  We now define a sesquibilinear 3-bracket $V \times V \times V \to V$, complex linear in the first two entries and complex antilinear in the third by
\begin{equation}
  \label{eq:3-bracket-C}
  [u,v;w]:= D(v,w) \cdot u~,
\end{equation}
for all $u,v,w \in V$.  The 3-bracket enjoys the following properties:
\begin{enumerate}
\renewcommand{\labelenumi}{(\alph{enumi})}
\item the \emph{unitarity} condition
  \begin{equation}
    \label{eq:unitarity-C}
    h([x,v;w],y) = h(x,[y,w;v])~;
  \end{equation}
\item the \emph{symmetry} condition
  \begin{equation}
    \label{eq:symmetry-C}
    h([x,v;w],y) = h([v,x;y],w)~;
  \end{equation}
\item and the \emph{fundamental identity}
  \begin{equation}
    \label{eq:FI-C}
    [[z,v;w],x;y] - [[z,x;y],v;w] - [z,[v,x;y];w] + [z,v;[w,y;x]] = 0~.
  \end{equation}
\end{enumerate}
The $N{=}6$ 3-algebra of \cite{BL4} obeys in addition the skewsymmetry condition
\begin{equation}
  \label{eq:skew-BL4}
  [x,y; z] = - [y,x; z]~.
\end{equation}
As proved in \cite[Theorem~22]{Lie3Algs}, this condition can be understood as the $111$ component of the Jacobi identity of a 3-graded complex Lie superalgebra $\fk = V \oplus \fg_\CC \oplus V^*$, in degrees $-1,0,1$ respectively.  The Lie bracket of $\fk$ extends the Lie bracket of $\fg_\CC$ and the actions of $\fg_\CC$ on $V$ and $V^*$ by defining the $[VV^*]$ bracket by $[v,\hfl{w}] = D(v,w)$.

\subsection{The quaternionic Faulkner construction}
\label{sec:quaternionic}

Although one could work quaternionically, we will think of quaternionic representations as complex representations with a quaternionic structure map.  Explicitly, by a quaternionic unitary representation of a Lie algebra $\fg$, we shall mean a complex unitary representation $V,h$ together with a $\fg$-invariant complex antilinear map $J:V \to V$, obeying $J^2 = -1$ and which is compatible with the hermitian structure $h$ in the sense that $h(Ju,Jv) = h(v,u)$.  Using $h$ and $J$ we may define a complex bilinear symplectic structure
\begin{equation}
  \label{eq:H-symplectic}
  \omega(u,v) = h(u,Jv)~.
\end{equation}
This allows us to transpose the $\fg$-action and define a \emph{bilinear} map $D_\omega : V \times V \to \fg_\CC$ by
\begin{equation}
  \label{eq:D-map-H}
  \left(D_\omega(u,v) , \XX\right) = \omega(\XX\cdot u, v)~,
\end{equation}
which in turn defines a \emph{trilinear} 3-bracket
\begin{equation}
  \label{eq:3-bracket-H}
  [u,v,w]:= D_\omega(u,v) \cdot w~.
\end{equation}
for all $u,v,w \in V$.  In terms of the sesquilinear map $D : V \times V \to \fg_\CC$ defined in \eqref{eq:D-map-C}, we have that $D_\omega(u,v) = D(u,Jv)$, whence $[u,v,w] = D_\omega(u,v) \cdot w = D(u, Jv) \cdot w = [w,u;Jv]$.  As shown in \cite{Lie3Algs}, the resulting complex triple system enjoys the following properties:
\begin{enumerate}
\renewcommand{\labelenumi}{(\alph{enumi})}
\item the \emph{symplecticity} condition
  \begin{equation}
    \label{eq:symplecticity-H}
    [u,v,w] = [v,u,w]~;
  \end{equation}
\item the \emph{symmetry} condition
  \begin{equation}
    \label{eq:symmetry-H}
    \omega([x, y, z ], w) = \omega([z , w, x], y)~,
  \end{equation}
\item the \emph{fundamental identity}
  \begin{equation}
    \label{eq:FI-H}
    [v, x, [w, y, z ]] − [w, y, [v, x, z ]] − [[v, x, w], y, z ] − [w, [v, x, y], z ] = 0~,
  \end{equation}
\item and the \emph{quaternionic} condition
  \begin{equation}
    \label{eq:quaternionic-H}
    J [x, y, z ] = [J x, J y, J z ]~.
  \end{equation}
\end{enumerate}
A particular case of these 3-algebras are those for which the 3-bracket obeys the cyclicity condition
\begin{equation}
  \label{eq:aLTS}
  [x,y,z] + [y,z,x] + [z,x,y] = 0~.
\end{equation}
Such 3-algebras are known as \textbf{anti-Lie triple systems}.  Anti-Lie triple systems admit an embedding in a complex Lie superalgebra in such a way that the 3-bracket is given by a nested Lie bracket.  Indeed, let $\fk = \fg_\CC \oplus V$ be 2-graded by declaring $\fg_\CC$ to have degree 0 and $V$ to have degree 1.
We give $\fk$ the structure of a Lie superalgebra by extending the Lie bracket on $\fg_\CC$ and the action of $\fg_\CC$ on $V$ as follows
\begin{equation}
  \label{eq:LSA}
  [v,w] = D_\omega(v,w) \in \fg_\CC~.
\end{equation}
Because of the symmetry of $D_\omega$, we see that this is indeed the Lie bracket of a Lie \emph{super}algebra.  The only component of the Jacobi identity which is in question is the $111$ component, but this is precisely the cyclicity condition \eqref{eq:aLTS}.

\section{Simplicity}
\label{sec:simplicity}

We have seen above that there are special cases of the Faulkner construction where the triple product is given by nested Lie brackets in an embedding graded Lie (super)algebra.  In the real case, the Lie triple systems embed in a 2-graded Lie algebra; in the complex case, the $N{=}6$ triple systems embed in a 3-graded Lie superalgebra; and in the quaternionic case, the anti-Lie triple systems embed in a Lie superalgebra.

In all these cases we have three different notions of simplicity and it makes sense that they should be related.  First we have a notion of simplicity of the triple system, by which we mean the absence of proper ideals.  Each triple system has an obvious notion of morphism and ideals are defined as kernels of morphisms.  We also have a notion of simplicity of the representation $V$, by which we mean irreducibility or, more precisely, indecomposability, with which it agrees in the case of positive-definite signature.  Finally we have a notion of simplicity of the embedding Lie (super)algebra, by which we mean again the absence of proper ideals.
As we will now show, in the case where the inner product on $V$ has positive-definite signature, these different notions of simplicity agree, with a minor exception in the case of the Lie triple systems (see below).

\subsection{Lie triple systems}
\label{sec:lie-triple-systems}

Let $V, \left<-,-\right>$ be a real faithful orthogonal representation of the real metric Lie algebra $\fg,\left(-,-\right)$.  Let $D: V \times V \to \fg$ be defined by \eqref{eq:D-map-R} and the 3-bracket by \eqref{eq:3-bracket-R}.  We will assume that the 3-bracket defines a Lie triple system, so that in addition to the unitarity \eqref{eq:unitarity-R}, symmetry \eqref{eq:symmetry-R} and fundamental identity \eqref{eq:FI-R}, it also obeys the cyclicity condition
\begin{equation}
  \label{eq:LTS}
  [x,y,z] + [y,z,x] + [z,x,y] = 0~.
\end{equation}
Let $\fk = \fg \oplus V$ be a 2-graded vector space with $\fg$ and $V$ having degrees $0$ and $1$, respectively.  We define on $\fk$ a 2-graded Lie algebra structure extending the Lie bracket on $\fg$ by declaring that, for all $x,y\in V$ and $X\in \fg$,
\begin{equation}
  \label{eq:LTS-2GLA}
  [x,y] = D(x,y) \qquad\text{and}\qquad [X,z] = X \cdot z~.
\end{equation}
The skewsymmetry condition \eqref{eq:D-skew} says that this is a Lie algebra and not a Lie superalgebra.  The Jacobi identities break up into the usual four homogeneous components: the $000$ component is zero by virtue of $\fg$ being a Lie algebra, the $001$ component by virtue of $V$ being a representation, the $011$ component by virtue of the fundamental identity \eqref{eq:FI-R}, rephrased as the $\fg$-equivariance of the map $D$, and the $111$ component is precisely the Lie triple system condition \eqref{eq:LTS}.  Furthermore $\fk$ is a metric 2-graded Lie algebra by declaring $\fg$ and $V$ to be perpendicular and using the given inner  products $(-,-)$ and $\left<-,-\right>$ on $\fg$ and $V$, respectively.  The ad-invariance of the inner product is built into the hypothesis of the construction.

We will assume that the inner product $\left<-,-\right>$ on $V$ is positive-definite.  This does not necessarily imply that the one on $\fg$ is positive-definite; although since $\fg$ is reductive, the inner product is arbitrary on the centre but a multiple (not necessarily of a definite sign) of the Killing forms of each of the simple ideals.

Let $V,W$ be two Lie triple systems and let $\varphi: V \to W$ be a morphism.  This means that for all $x,y,z \in V$, $[\varphi x, \varphi y, \varphi z] = \varphi[x,y,z]$.  Then $U = \ker \varphi$ is an \textbf{ideal} of the Lie triple system, which means that
\begin{equation}
  \label{eq:LTS-ideal}
  [UVV]\subset U \qquad\text{and}\qquad [VVU] \subset U~,
\end{equation}
in the obvious notation.

\begin{theorem}
  \label{thm:LTS-simplicity}
  Let $V$ be a positive-definite metric Lie triple system, let $\fg$ be the image of the Faulkner map $D$ and let $\fk = \fg \oplus V$ denote its embedding 2-graded Lie algebra.  The following are equivalent:
  \begin{enumerate}
  \item $V$ is irreducible as a $\fg$-module,
  \item $V$ is simple (or one-dimensional) as a Lie triple system,
  \item $\fk$ is a simple (or one-dimensional) 2-graded Lie algebra.
  \end{enumerate}
\end{theorem}

\begin{proof}
  The second and third conditions can be paraphrased as saying that $V$ has no proper ideals and $\fk$ has no proper \emph{homogeneous} ideals, since this is what it means to be simple as a \emph{2-graded} Lie algebra.  This means that they are simple or one-dimensional.  We will prove the contrapositive of each of the statements except the last, which shall be proved directly.
  
  \emph{$(1) \implies (2)$}.  Suppose that $U \lhd V$ is a proper ideal of the Lie triple system.  Then the second equation in \eqref{eq:LTS-ideal} says that $U$ is a proper $\fg$-submodule of $V$.

  \emph{$(3) \implies (1)$}.  Suppose that $W < V$ is a proper $\fg$-submodule of $V$, then $I = [WW]\oplus W$ is a proper ideal of $\fk$.  Indeed, we first notice that by the positive-definiteness assumption $V = W \oplus W^\perp$, for $W^\perp$ the perpendicular complement.  We then observe that $[WW^\perp] = 0$, since for all $w \in W$, $x \in W^\perp$ and $X \in \fg$, $\left([wx],X\right) = \left<X\cdot w, x\right>$, but $X \cdot w \in W$ since it is a submodule and hence the inner product with $x$ is zero.  This says that $[W^\perp I] = 0$.  By construction $[W I] \subset I$ and because $W$ is a $\fg$-submodule, $[\fg I] \subset I$.

    \emph{$(2) \implies (1)$}.  Suppose again that $W<V$ is a proper $\fg$-submodule of $V$.  Then $W\lhd V$ is a proper ideal of $V$ as a Lie triple system.  For this we show that $[VVW] \subset W$ and $[WVV] \subset W$.  The former is precisely the fact that $W<V$ is a $\fg$-submodule.  For the latter, notice that $[WW^\perp V] =0$ because $[WW^\perp]=0$, which we proved above.  Therefore $[WVV]=[WWV]$.  Now by cyclicity, $[WWV] \subset [VWW] \subset W$, since $W<V$ is a $\fg$-submodule.

    \emph{$(1) \implies (3)$}.  Let $I = I_0 \oplus I_1 \lhd \fk$ be a homogeneous ideal.  Then $[\fg I_1] \subset I_1$ says that $I_1 \subset V$ is an $\fg$-submodule.  Since $V$ is irreducible, $I_1=0$ or $I_1 = V$.  In the former case, $I = I_0$ and $[I_0 V] = 0$, but $\fg$ acts faithfully on $V$, whence $I_0=0$ and hence $I=0$.  In the latter case, $I_1=V$ then $[VV]\subset I_0$, but $[VV]=\fg$, whence $I_0=\fg$ and $I=\fk$. So $\fk$ has no proper homogeneous ideals.
\end{proof}

The theorem can be strengthened by substituting condition $(3)$ in the statement of the theorem with
\begin{itemize}
\item[$(3)'$] $\fk$ is a simple (or one-dimensional) Lie algebra or else $V \cong \fg$ (as $\fg$-modules) and $\fg$ is a simple (or one-dimensional) Lie algebra.
\end{itemize}
The distinction is that in the case where $V \cong \fg$, the diagonal embedding $\fg \into \fg \oplus \fg$ defines a non-homogeneous proper ideal of $\fk$.  The embedding Lie algebra $\fg$ is then rendered isomorphic to $\fg \oplus \fg$ as a Lie algebra without 2-grading, with the ideal corresponding to one of the two copies of $\fg$.

\begin{proof}[Proof that $(1) \implies (3)'$]
  We notice that by the theorem it is enough to show that an ideal is homogeneous, to conclude that it cannot be proper.  This will be used implicitly in order to simplify the exposition.  Let $I \lhd \fk$ be an ideal, not necessarily homogeneous.  Let $W = I \cap V$.  It is a $\fg$-submodule of $V$, whence it is either $0$ or $V$, since $V$ is irreducible.  If $W = V$, then $I = I_0 \oplus V$ is homogeneous.  If $W=0$ then $I$ is transversal to $V$.  Let $\bar I$ denote the projection of $I$ onto $V$.  The projection is $\fg$-equivariant, hence $\bar I < V$ is a $\fg$-submodule.  Again irreducibility says that either $\bar I=0$, in which case $I \subset \fg$ is homogeneous, or else $\bar I=V$.  In this latter case, $I = I_0 \oplus \Gamma$, where $\Gamma$ is the graph of a linear map $\varphi: V \to \fg$.  Since $I$ is an ideal, in particular $[\fg I] \subset I$, whence in particular $[\fg I_0] \subset I_0$, and hence $I_0$ is an ideal of $\fg$.  Since $\fg$ is reductive, there's a complementary ideal $I_0^\perp$ and by suitably redefining $\varphi$, we can assume that $\varphi: V \to I_0^\perp$.  Again since $I$ is an ideal, in particular $[\fg \Gamma] \subset \Gamma$, which is equivalent to the $\fg$-equivariance of $\varphi$.  This means that $\ker\varphi < V$ is a $\fg$-submodule.  Again irreducibility says that either $\ker\varphi = V$, in which case $\varphi = 0$ and hence $\Gamma =0$, whence $I = I_0$ is homogeneous, or else $\ker\varphi =0$ and $\varphi$ is one-to-one.  Its image is thus a $\fg$-submodule of $I_0^\perp$; that is, an ideal of $\fg$.  Since $\fg$ is reductive, we have $I_0^\perp = \im \varphi \oplus J_0$, for some ideal $J_0 \lhd \fg$.  In summary, $I = I_0 \oplus \Gamma$ and $\fg = I_0 \oplus \im \varphi \oplus J_0$.  Now, since $I$ is an ideal, $[VI] \subset I$, which in particular implies that $[V I_0] \subset I \cap V = 0$.  Since $V$ is a faithful $\fg$-module, this means $I_0=0$.  Similarly, $[VI]\subset \Gamma$, which unpacks to the following: that for all $v,w\in V$,
  \begin{equation*}
    [v, w + \varphi w] = [v,w] + [v,\varphi(w)] \in \Gamma \implies [v,w] = \varphi[v,\varphi w]~,
  \end{equation*}
which says that $\fg = \im\varphi$, or that $J_0 = 0$.  In other words, $\varphi: V \to \fg$ is an isomorphism of $\fg$-modules.  Since $V$ is irreducible, $\fg$ is irreducible as an adjoint module, whence $\fg$ has no proper ideals: it is thus simple or one-dimensional.  In summary, $\fk = \fg_0 \oplus \fg_1$, as a vector space, where both $\fg_0$ and $\fg_1$ are isomorphic to $\fg$, with subscripts indicating the degrees.  The Lie bracket is 2-graded.  Explicitly, if we let $X_a$, for $a=0,1$, denote the image of $X\in \fg$ in $\fg_a$, then $[X_a,Y_b] = [X,Y]_{a+b}$, where subscript addition is taken modulo $2$.  The ideal $I = \left\{X_0 + X_1\middle | X \in \fg\right\}$ is the diagonal embedding of $\fg$.  Ignoring the 2-grading, $\fk = I \oplus J$, where both $I$ and $J= \left\{X_0 - X_1 \middle | X \in \fg\right\}$ are now commuting ideals isomorphic to $\fg$ as Lie algebras.
\end{proof}

Of course these results are classical.  They follow from the classification of irreducible riemannian symmetric spaces, which is equivalent to the classification of simple, positive-definite Lie triple systems.  We include them here because of completeness and in order to illustrate in a simpler case the less familiar complex and quaternionic cases below.

\subsection{N=6 triple systems}
\label{sec:n=6-triple-systems}

Let $V,h$ be a faithful complex unitary representation of a real metric Lie algebra $\fg,\left(-,-\right)$ giving rise to a sesquilinear Faulkner map $D: V \times V \to \fg_\CC$ as in \eqref{eq:D-map-C} and hence to 3-bracket $[x,y;z] := D(y,z) \cdot x$ as in \eqref{eq:3-bracket-C}.  Assume furthermore that in addition to the unitarity \eqref{eq:unitarity-C}, symmetry \eqref{eq:symmetry-C} and fundamental \eqref{eq:FI-C} identities, it also satisfies the skewsymmetry condition \eqref{eq:skew-BL4}.  We call such 3-algebras an \textbf{N=6 triple system}.  Similarly to the case of Lie triple systems, an N=6 triple system embeds in a 3-graded Lie superalgebra in such a way that the 3-bracket is giving by nesting the Lie bracket.  Indeed, on the complex vector space $\fk = V \oplus \fg_\CC \oplus V^*$ the Lie bracket on $\fg_\CC$ together with
\begin{equation}
  \label{eq:3G-LSA}
  [u,\hfl{v}] = D(u,v), \qquad  [\XX,u] = \XX \cdot u \qquad\text{and}\qquad [\XX,\hfl{v}] = \hfl{\overline \XX \cdot v}
\end{equation}
define the structure of a 3-graded Lie superalgebra, with $V,\fg_\CC,V^*$ having degrees $-1,0,1$ respectively.  The Jacobi identity breaks up into 5 homogeneous components: $-1-11$, $-100$, $-101$, $000$, $001$, $-111$, all other components being zero automatically because $\fk$ has no elements of degree $d$ with $|d|>1$.  The $000$ component is the Jacobi identity for $\fg_\CC$, the $001$ and $-100$ components vanish by virtue of $V^*$ and $V$ being $\fg_\CC$-modules and the $-101$ component vanishes because $D$ is $\fg_\CC$-equivariant, which is the content of the fundamental identity.  The remaining components vanish by the skewsymmetry condition \eqref{eq:skew-BL4}, whose sign is responsible for the correct Jacobi identity for a Lie \emph{super}algebra, as opposed to a Lie algebra.  Furthermore $\fk$ admits an ad-invariant complex bilinear inner product extending the one on $\fg_\CC$ by declaring $\fg_\CC$ perpendicular  to $V \oplus V^*$ and defining
\begin{equation}
  \label{eq:3GLSA-IP}
  (u,\hfl{v}) = h(u,v) = - (\hfl{v},u)~.
\end{equation}
The ad-invariance of this inner product is built into the Faulkner construction, as explained in \cite{Lie3Algs}.  We will assume in what follows that $h$ is positive-definite.

A morphism $\varphi: V \to W$ of N=6 triple systems is a complex linear map obeying $[\varphi u, \varphi v ; \varphi w] = \varphi[u, v; w]$ and $h(\varphi u, \varphi v) = h(u,v)$; although this latter identity plays no rôle in the following definition.  The kernel of $\varphi$ defines the notion of an \textbf{ideal} for an N=6 triple system.  In other words, an ideal is a complex subspace $U < V$ such that
\begin{equation}
  \label{eq:BL4-ideal}
  [UVV] \subset U \qquad\text{and}\qquad [VVU] \subset U~.
\end{equation}
The first condition says that $U$ is a $\fg_\CC$-submodule of $V$, whereas in positive-definite signature, the second condition is a consequence of the first as we will see below.

Analogously to Theorem~\ref{thm:LTS-simplicity}, we have the following result, a version of which is stated in \cite{Palmkvist}.

\begin{theorem}
  \label{thm:BL4-simplicity}
  Let $V$ be a positive-definite N=6 triple system, let $\fg$ be the Faulkner Lie algebra and let $\fk = V \oplus \fg_\CC \oplus V^*$ denote its embedding 3-graded Lie superalgebra.  The following are equivalent:
  \begin{enumerate}
  \item $V$ is irreducible as a $\fg$-module,
  \item $V$ is simple (or one-dimensional) as an N=6 triple system,
  \item $\fk$ is a simple (or one-dimensional) 3-graded Lie superalgebra.
  \end{enumerate}
\end{theorem}

\begin{proof}
  Again, the second and third conditions can be paraphrased as saying that $V$ has no proper ideals and $\fk$ has no proper homogeneous ideals, since this is what it means to be simple as a \emph{3-graded} Lie superalgebra.  This means that they are simple or one-dimensional.  We will prove the contrapositive of each of the statements except the last.
  
  \emph{$(2) \implies (1)$}.  Suppose that $V$ is reducible as a $\fg$-module and let $U < V$ be a proper $\fg$-submodule.  Its perpendicular complement
  \begin{equation*}
    U^\perp = \left\{v \in V \middle | h(u,v)=0\,\forall u \in U\right\}
  \end{equation*}
  relative to the positive-definite hermitian inner product is also a submodule and $V = U \oplus U^\perp$.  Then this is also a decomposition of the triple system $V$ into as a direct sum of perpendicular ideals.  The first observation is that $D(U,U^\perp)=0$, in the obvious notation.  Indeed, let $u \in U$ and $v \in U^\perp$, then for any $\XX \in \fg_\CC$, we have $\left(D(u,v),\XX\right) = h(\XX \cdot u, v)$.  But, since $U$ is a submodule, $\XX \cdot u \in U$ and hence is perpendicular to $v$.  This means that $[VUU^\perp] = 0$ and also by complex conjugation, $[VU^\perp U]=0$.  Using the skewsymmetry condition \eqref{eq:skew-BL4}, it also means that $[U^\perp VU]=0$ and $[UVU^\perp]=0$.  Hence the only nonzero 3-brackets are $[UUU]$ and $[U^\perp U^\perp U^\perp]$.  Furthermore $[UUU] \subset U$, since $U$ is a submodule, and similarly for $U^\perp$.

  \emph{$(1) \implies (2)$}.  Let $V = U \oplus U^\perp$ be a decomposition of the triple system $V$ into perpendicular proper ideals.  This means that the only 3-brackets which are nonzero are $[UUU]\subset U$ and $[U^\perp U^\perp U^\perp]\subset U^\perp$.  In particular, $[UVV]\subset U$, which says that $U<V$ is a proper submodule.

  \emph{$(3) \implies (1)$}.  As above let $V = U \oplus U^\perp$ be a reducible $\fg$-module with proper submodules $U$ and $U^\perp$.  Define
  \begin{equation*}
    U^* = \{ \hfl{u} \mid u \in U \} < V^*~,
  \end{equation*}
 which obeys $(U^*)^\perp = (U^\perp)^*$, where this latter space is defined in a similar way.  In terms of the embedding Lie superalgebra, the vanishing of the 3-brackets above mean that $[U(U^\perp)^*]=0$ and $[U^\perp U^*] = 0$ and also that $[[UU^*]U^\perp]$, $[[UU^*](U^\perp)^*]$, $[[U^\perp (U^\perp)^*]U]$ and $[[U^\perp (U^\perp)^*]U^*]$ all vanish.  This then implies that $U \oplus [UU^*] \oplus U^*$ is a homogeneous ideal of $\fk$.  In fact, its perpendicular complement in $\fk$ is 
$U^\perp \oplus [U^\perp (U^\perp)^*] \oplus (U^\perp)^*$.

  \emph{$(1) \implies (3)$}.  Let $I = I_{-1} \oplus I_0 \oplus I_1 \lhd \fk$ be a homogeneous ideal.  Then in particular $[\fg_{\CC}I]\subset I$, whence $I_{\pm 1}$ are submodules of $V$ and $V^*$, respectively.  Since $V$ and $V^*$ are irreducible, this means that either $I_1 = 0$ or $I_1 = V^*$ and similarly that either $I_{-1}=0$ or $I_{-1}=V$.  Thus there are four cases to consider:
  \begin{enumerate}
    \renewcommand{\labelenumi}{(\roman{enumi})}
  \item $I=I_0$.  Then $[VI]= 0$ since $I$ has no piece of degree $-1$, but since $V$ is a faithful representation, $I=0$.
  \item $I=I_0 \oplus V^*$.  Again $[VI_0] =0$ and by faithfulness $I_0=0$, so that $I = V^*$, but this is not an ideal because $[VV*]\neq 0$.
  \item $I=V \oplus I_0$.  Now $[V^* I_0] =0$ since $I$ has no piece of degree $1$.  Since $\fg_\CC$ acts faithfully on $V^*$, this means that $I_0=0$, whence $I=V$, which is not an ideal.
  \item $I= V \oplus I_0 \oplus V^*$.  $[VV*]\subset I_0$ since $I$ is an ideal, but also $[VV*]=\fg_\CC$, whence $I_0=\fg_\CC$ and $I=\fk$.
  \end{enumerate}
  In summary, $\fk$ has no proper homogeneous ideals.
\end{proof}

As in the case of Lie triple systems, we may strengthen the theorem by substituting 
condition $(3)$ in the statement of the theorem with
\begin{itemize}
\item[$(3)'$] $\fk$ is a simple (or one-dimensional) Lie superalgebra.
\end{itemize}
In other words, in this case $\fk$ has no proper ideals of any kind.

\begin{proof}[Proof that $(1) \implies (3)'$]
  Again, the theorem allows us to discard any ideals which are homogeneous, since they cannot be proper.  Let $I = I_0 \oplus J$ be an ideal of $\fk$, with $I_0 \subset \fg_\CC$ and $J \subset V \oplus V^*$.  Then $I_0 \lhd \fg_\CC$ is an ideal and $J \subset V \oplus V^*$ is an $\fg_{\CC}$-submodule, whence so are $J \cap V$ and $J \cap V^*$.  Since $V$ (and hence $V^*$) are irreducible, these cannot be proper submodules.  This means that we have again four cases to treat:
  \begin{enumerate}
    \renewcommand{\labelenumi}{(\roman{enumi})}
  \item $J \cap V = V$ and $J \cap V^* = V^*$.  Then $J = V \oplus V^*$, so that $I$ is homogeneous.
  \item $J \cap V = V$ but $J \cap V^* = 0$.  Then $J=V$ and again $I$ is homogeneous.
  \item $J \cap V = 0$ but $J \cap V^* = V^*$.  Then $J=V^*$ and again $I$ is homogeneous.
  \item $J \cap V = 0$ and $J \cap V^* = 0$.  This means that $J$ is transversal to both $V$ and $V^*$.  Let $\pi: J \to V$ denote the projection of $J$ onto $V$ along $V^*$.  Since $\pi$ is $\fg_\CC$-equivariant, the image $\pi(J) \subset V$ is a submodule.  Hence it is either $0$ or $V$.  If $\pi(J)=0$, then $J \subset V^*$, but since $J\cap V^* = 0$, this shows that $J=0$ and $I$ is again homogeneous.  Finally, if $\pi(J) = V$, then $J$, being a $\fg_\CC$-submodule is the graph
    \begin{equation*}
      J = \left\{v + \varphi v \middle |  v \in V\right\}
    \end{equation*}
    of a $\fg_\CC$-equivariant linear map $\varphi: V \to V^*$.  Since $I$ is an ideal, in particular $[V^* I]\subset I$, whence $[V^* J] \subset I_0$, but $[V^* J] = [V^*V] = \fg_\CC$, whence $I_0 = \fg_\CC$.  But then the ideal condition implies that $[V\oplus V^*, I_0] \subset J$, which is absurd since $[V \oplus V^*, \fg_\CC] = V \oplus V^*$.  Therefore there is no such ideal.
  \end{enumerate}
\end{proof}

\subsection{Quaternionic anti-Lie triple systems}
\label{sec:n=4-5-triple}

Let $V,h,J$ be a quaternionic anti-Lie triple system.  Recall that we think of it as a complex anti-Lie triple system with a compatible quaternionic structure.  This means that the complex trilinear 3-bracket on $V$ satisfies the conditions \eqref{eq:symplecticity-H}, \eqref{eq:symmetry-H}, \eqref{eq:FI-H}, \eqref{eq:quaternionic-H} and \eqref{eq:aLTS}, where $\omega$ is the complex symplectic structure defined by $h$ and $J$ by \eqref{eq:H-symplectic}.  As explained in Section~\ref{sec:quaternionic}, we may embed the anti-Lie triple system into a metric complex Lie superalgebra $\fk = \fg_\CC \oplus V$, where $\fg_\CC$ is the complexification of the Faulkner Lie algebra $\fg$.  In this section we will assume that the hermitian structure $h$ is positive-definite.

Let $\varphi: V \to W$ be a morphism of quaternionic anti-Lie triple systems.  This means that $\varphi$ is a complex linear map, commuting with $J$, and preserving the 3-bracket: $\varphi[u,v,w] = [\varphi u, \varphi v, \varphi w]$.  Then the kernel of $\varphi$ is an ideal and all ideals are of this form.  This means that an ideal $U \subset V$ is a complex subspace, stable under $J$, and obeying
\begin{equation}
  \label{eq:aLTS-ideal}
  [VVU] \subset U \qquad\text{and}\qquad [UVV] \subset U~.
\end{equation}
The first of the above equations says that $U < V$ is a quaternionic $\fg$-submodule of $V$.

In dealing with quaternionic representations, we have to distinguish between two notions of irreducibility: irreducibility as a quaternionic representation or irreducibility as a complex representation.  We shall refer to the former kind as \emph{quaternionic-irreducible} representations.  They are characterised by the fact that they admit no proper quaternionic submodules; that is, submodules stable under $J$.  A quaternionic-irreducible representation need not be irreducible, but it is easy to characterise those which are not.

\begin{lemma}
  Let $V$ be a quaternionic-irreducible $\fg$-module.  Then either $V$ is irreducible or else $V = W \oplus JW$, where $W$ is irreducible.
\end{lemma}

\begin{proof}
  By hypothesis $V$ admits no proper quaternionic submodules.  Then either $V$ is irreducible or else it admits a proper submodule $W<V$, which is not stable under the action of the quaternionic structure map $J$.  By the invariance of $J$, $JW$ is also a submodule of $V$.  The linear span $W + JW$ of the subspaces $W$ and $JW$ is stable under $J$, whence it is a quaternionic submodule.  Since it cannot be proper it is all of $V$, since $W \neq 0$.  The intersection $W \cap JW$ is also a quaternionic submodule of $V$, whence again not proper.  It cannot be all of $V$ (since $W$ is proper), so it must be $0$.  Therefore $V = W \oplus JW$.  Finally notice that $W$ itself is irreducible, for if $U < W$ is a proper submodule, then $U \oplus JU$ would be a proper quaternionic submodule of $V$, which contradicts the hypothesis.
\end{proof}

\begin{theorem}
  \label{thm:aLTS-simplicity}
  Let $V$ be a positive-definite quaternionic anti-Lie triple system, let $\fg$ be the corresponding Faulkner Lie algebra and let $\fk = \fg_\CC \oplus V$ denote its embedding Lie superalgebra.  The following are equivalent:
  \begin{enumerate}
  \item $V$ is a quaternionic-irreducible $\fg$-module,
  \item $V$ is a simple quaternionic anti-Lie triple system,
  \item $\fk$ is a simple Lie superalgebra.
  \end{enumerate}
\end{theorem}

Let us remark that $V$, being quaternionic, cannot be one-dimensional, hence neither is $\fk$.  This means that the absence of proper ideals does imply simplicity.

\begin{proof}
  We will prove the contrapositive of each of the statements, except the last.

  \emph{$(3) \implies (1)$}.  Suppose that $V= W \oplus W^\perp$ is reducible as $\fg$-module, where $W$ is a proper submodule and $W^\perp$ is its perpendicular complement.  It does not matter whether we define perpendicularity using $h$ or $\omega$, since $W$ is quaternionic by assumption and hence stable under $J$, hence the two notions agree.  This implies that $D_\omega(W,W^\perp) = 0$, in the obvious notation.  Indeed, if $w \in W$ and $v \in W^\perp$, then for all $\XX \in \fg_\CC$, $(D_\omega(w,v),\XX) = \omega(\XX \cdot w, v)$.  Now, since $W$ is a submodule, $\XX \cdot w \in W$, which is then perpendicular to $v \in W^\perp$.  In the embedding Lie superalgebra, this means that $[WW^\perp] = 0$, which in turn implies that $[WW]\oplus W$ is a proper ideal of $\fk$.  Indeed, since $W$ is a submodule, $[\fg_\CC W] \subset W$ and hence $[\fg_\CC [WW]] \subset [WW]$.  Similarly, $[VW] \subset [WW]$, since $[W^\perp W] = 0$.  Finally $[V[WW]] \subset [[VW]W] \subset W$, since $W$ is a submodule.

  \emph{$(2) \implies (1)$}.  Again let $V =  W \oplus W^\perp$, with $W$ and $W^\perp$ as above.  Then $[VVW] \subset W$, since $W$ is a submodule.  Now notice that $D_\omega(W,W^\perp) = 0$ implies that $[WW^\perp V] =0$, whence $[WVV] = [WWV]$.  Now by cyclicity, $[WWV] \subset [VWW] \subset W$.  Therefore $W$ is a proper ideal of $V$ as an anti-Lie triple system.

  \emph{$(1) \implies (2)$}.  Let $V = W \oplus W^\perp$ be a decomposition of $V$ as an anti-Lie triple system into perpendicular proper ideals.  Then in particular, $[VVW] \subset W$ and hence $W$ is a proper submodule.

  \emph{$(1) \implies (3)$}.  Let $V$ be a quaternionic-irreducible module and suppose that $I = I_0 \oplus I_1$ is an ideal of the Lie superalgebra $\fk$.  Then $[\fg_\CC I_1] \subset I_1$ says that $I_1 < V$ is a submodule.  By the lemma, either $V$ is itself irreducible, or else $V = W \oplus JW$, where $W$ is irreducible.  In this latter case, positive-definiteness of the inner product says that $JW = W^\perp$, so that $W$ and $JW$ are lagrangian submodules with respect to the $\omega$, whence $[W,W]=0=[JW,JW]$ in $\fk$.
  \begin{enumerate}
    \renewcommand{\labelenumi}{(\roman{enumi})}
  \item \emph{$V$ is irreducible.}  In this case, $I_1= 0$ or else $I_1 = V$.  If $I_1=0$, then $I=I_0$.  But then $[VI_0]=0$, but since $\fg_\CC$ acts faithfully on $V$, $I_0=0$, whence $I=0$.  Alternatively, if $I_1=V$, then $I= I_0 \oplus V$.  But $[VV] \subset I_0$ says that $I_0 = \fg_\CC$, whence $I= \fk$.  In summary, $\fk$ has no proper ideals.
  \item \emph{$V = W \oplus JW$, with $W$ irreducible.}  In this case, $I_1$ can be one of four possibilities: $0$, $W$, $JW$, $V$.  The first and last suffer the same fate as above: $I$ is not a proper ideal.  The case $I_1=W$ and $I_1=JW$ are similar, so we treat only the first.  If $I_1 = W$, then $[VI_1] \subset I_0$, but $[VW] = [JW,W] = [VV] = \fg_\CC$, whence $I_0=\fg_\CC$.  But then $[JW,I_0] \subset I_1 = W$, which is absurd since $JW$ is a submodule.  Hence there are no such ideals.
  \end{enumerate}
\end{proof}

\subsection{Classifications}
\label{sec:classifications}

The above theorems reduce the classification of positive-definite simple Lie, N=6, and quaternionic anti-Lie triple systems to the well-known classification of simple Lie (super)algebras.  None of these classifications are new, of course.  In the positive-definite case, the classification of simple Lie triple systems is precisely the classification of irreducible riemannian symmetric spaces, which goes back to Élie Cartan and is described, for example, in \cite{Helgason}.  The classification of simple Lie triple systems, without any metricity assumption, is due to Lister \cite{Lister}.  The classification of simple positive-definite N=6 triple systems follows from the classification of simple Lie superalgebras, which is sketched in \cite{KacSuperSketch} and perhaps more accessibly also in \cite{DictLieSuperAlg}.  As recapitulated in \cite{Palmkvist} and mentioned already in \cite{Lie3Algs}, the corresponding simple Lie superalgebras are $A(n,m)$ and $C(n+1)$, agreeing with the classification in \cite{SchnablTachikawa} and remarks in \cite{3Lee,GaiottoWitten}.  Finally, the simple positive-definite quaternionic anti-Lie triple systems are in one-to-one correspondence with the simple Lie superalgebras whose fermionic subspace admits a quaternionic structure.  These are given by the basic classical Lie superalgebras $A(m,n)$, $C(n+1)$, $B(m,n)$, $D(m,n)$, $F(4)$, $G(3)$ and $D(2,1;\alpha)$, in agreement with the results of \cite{BHRSS} obtained via conformal limits of gauged three-dimensional supergravities.

\section*{Acknowledgments}

This work was supported by World Premier International Research Center Initiative (WPI Initiative), MEXT, Japan.  It was done while on sabbatical at the Institute for the Physics and Mathematics of the Universe (IPMU) at the University of Tokyo, whose hospitality and support is hereby acknowledged.  It is a pleasure to thank IPMU's director Hitoshi Murayama who made this visit possible as well as the Leverhulme Trust for the award of a Research Fellowship freeing me from my teaching and administrative duties at the University of Edinburgh.  Last, but not least, I would like to thank my collaborators Paul de Medeiros and Elena Méndez-Escobar for the continuing 3-algebraic discussions.  In particular, I am thankful to Paul for a question which revealed an error in the proof of Theorem \ref{thm:aLTS-simplicity} in a previous version of this paper.

\bibliographystyle{utphys}
\bibliography{AdS,AdS3,ESYM,Sugra,Geometry,Algebra}

\end{document}